\documentclass[conference]{IEEEtran}
\IEEEoverridecommandlockouts
\usepackage{cite}
\usepackage{amsmath,amssymb,amsfonts}
\usepackage{algorithmic}
\usepackage{textcomp}
\usepackage{xcolor}
\def\BibTeX{{\rm B\kern-.05em{\sc i\kern-.025em b}\kern-.08em
    T\kern-.1667em\lower.7ex\hbox{E}\kern-.125emX}}

\usepackage{graphicx}
\usepackage{algorithm}
\usepackage{algorithmic}
\usepackage{amsmath}
\usepackage{bm}
\usepackage{indentfirst}
\usepackage{amsthm}
\usepackage{multicol}
\usepackage{multirow}
\usepackage{color}
\usepackage{subfigure}

\usepackage{authblk}
\usepackage{threeparttable}
\usepackage{listings}
\usepackage[numbers,sort&compress]{natbib}
\usepackage{longtable}
\usepackage{rotating}
\usepackage{stfloats}
\usepackage[colorlinks=true,linkcolor=black,citecolor=blue,urlcolor=blue]{hyperref}
\usepackage[doipre={doi:~}]{uri}
\usepackage{url}

\usepackage{mathrsfs}
\usepackage{amsmath}
\usepackage{booktabs}
\usepackage{soul}
    \ifCLASSINFOpdf
  \graphicspath{{Figures}{}}
\else
  \graphicspath{{../Figures}}
\fi

\begin{document}

\title{EHAP-ORAM: Efficient Hardware-Assisted Persistent ORAM System for Non-volatile Memory}

\author[a, b]{Gang Liu}
\author[a]{Kenli Li}
\author[a]{Zheng Xiao}
\author[b]{Rujia Wang}
\affil[a]{College of Information Science and Engineering, Hunan University}
\affil[b]{Computer Science Department, Illinois Institute of Technology}
\affil[ ]{\{liug, lkl, zxiao\}@hnu.edu.cn and  rwang67@iit.edu}

\maketitle

\begin{abstract}

Oblivious RAM (ORAM) is a provable secure primitive to prevent access pattern leakage on the memory bus. It serves as the intermediate layer between the trusted on-chip components and the untrusted external memory systems to modulate the original memory access patterns into indistinguishable memory sequences.
By randomly remapping the data blocks and accessing redundant blocks, ORAM prevents access pattern leakage through obfuscation. While there is much prior work focusing on improving ORAM's performance on the conventional DRAM-based memory system, when the memory technology shifts to use non-volatile memory (NVM), new challenges come up as how to efficiently support crash consistency for ORAM.

In this work, we propose EHAP-ORAM, which studies how to persist ORAM construction with an NVM based memory system.
We first analyze the design requirements for a persistent ORAM system and discuss the need to preserve crash consistency and atomicity for both data and ORAM metadata.
Next, we discuss some of the challenges in the design of a persistent ORAM system and propose some solutions to those challenges.
Then, we propose the modified on-chip ORAM controller architecture.
Based on the improved hardware architecture of the ORAM controller on-chip, we propose different persistency protocols to ensure the crash consistency of the ORAM system and satisfy that the metadata in PosMap is safe when it is persisted to NVM in trusted/untrusted off-chip.
The proposed architecture and persistency protocol steps minimize the overhead and leakage during the write-back process.
Finally, we compared our persistent ORAM with the system without crash consistency support, show that in non-recursive and recursive cases, EHAP-ORAM only incurs 3.36\% and 3.65\% performance overhead.
The results show that the EHAP-ORAM can support efficient crash consistency with minimal performance and hardware overhead.

\end{abstract}



\section{Introduction}
\label{s:intro}

Protecting the security and privacy of the data and program running on a shared system is never easy. There is an increasing need for system designers to consider security and privacy protection in addition to performance. There are a lot of efforts from the industry and academia designing secure hardware to give the system a root-of-trust. For example, TPM \cite{bajikar2002trusted}, SGX \cite{johnson2016intel}, XOM \cite{lie2000architectural}, Trustzone \cite{trustzone} and SME \cite{kaplan2016amd}, process sensitive data through data encryption and integrity check, or reserve a protected region that cannot be tampered, which effectively prevent adversaries from revealing the plaintext or compromising the data easily. However, the protections are still mainly using encryption and integrity check, which is far from enough. Because of the sharing, malicious applications are able to probe sensitive information from victim applications through various side channels. For instance, the timing information, the power usage and the memory access pattern can be exploited by malicious adversaries to infer sensitive information. Among them, memory access pattern leakage refers to that the adversaries can utilize the temporal and spatial information on the memory address bus to correlate the program's control flow graph \cite{zhuang2004hide}, the searchable encryption database \cite{islam2012access}, or even the neural network structure \cite{hu2020deepsniffer}.


The cryptographic community proposed Oblivious RAM (ORAM) \cite{goldreich1987towards, goldreich1996software} to address the memory access pattern leakage. The ultimate goal of ORAM is to hide the program access pattern by adding redundant blocks and periodically reshuffling the data in memory. In this way, the attacker will be not able to guess whether the program is accessing the same or a different data, whether the access is a read or a write, whether we are repeatedly accessing a hot region, etc. The efficiency of ORAM family has improved significantly in recent years. Tree-based ORAM, such as Path ORAM \cite{stefanov2013path}, has become one of the mainstream ORAM protocols that people adopt to use on main memory systems\cite{sasy2017zerotrace,fletcher2015low}, with an access overhead of $log(N)$ ($N$ is the number of total blocks in the memory).

The computer architecture community has started to optimize ORAM with extensive hardware and software co-design, mainly focus on performance \cite{ren2013design,zhang2015fork,wang2018d}. Also, most of these works assume that main memory uses the mature DRAM technology. As DRAM faces the inevitable scaling challenges, more and more vendors are investing in emerging non-volatile memory (NVM) technology. For example, 3dXpoint based Optane memory \cite{hady2017platform} has already been released to the public; future computing systems such as memory-centric computing architecture \cite{keeton2015machine,benton2017ccix},  these methods prefer to use NVM as the unified memory backend.
Compared to DRAM, NVM provides natural benefits such as non-volatility, persistency, and high-density, which makes it to be the future memory system’s candidate.
On the other hand, NVM based memory system still faces the same security issue: the memory address bus can leak information through the access pattern. While some prior work start to address at this issue \cite{awad2017obfus} through trusted components on the NVM DIMM, such a threat model is very strong, as the security boundary now include components that are off-chip. The only provable secure way to defend the memory system from access pattern leakage is to utilize ORAM, and implementing ORAM on NVM brings us more challenging questions to be answered.

{When implementing the ORAM protocol with NVM systems, it still suffers from crash consistency issues, just like other secure memory systems with encryption or integrity check \cite{yang2019no, liu2018crash}.
The specific requirement to address crash consistency is that application data (e.g., documentation, data, and configuration) must be recoverable even if the system power fails or the system crashes \cite{jiang2016crash, pillai2014all}.
Traditional software-based solutions, such as logging \cite{coburn2011nv, volos2011mnemosyne} or copy-on-write mechanism (CoW) \cite{condit2009better, venkataraman2011consistent}, can handle data recovery well.
However, such approaches cannot work well with NVM-ORAM systems for two reasons.
First, software-based (e.g., logging or copy-on-write) support for crash consistency mechanisms is inefficient \cite{liu2018crash, ren2015thynvm}.
Second, it may lead to information leakage and break ORAM system protection.
The analysis of these two reasons is in Section \ref{SoftwareBasedInefficientandInsecurity}.}

{In this paper, we would like to study the crash consistency problem when we implement ORAM protocols with the NVM system.
By improving the ORAM hardware architecture and software protocol, we propose an end-to-end EHAP-ORAM architecture. }
EHAP-ORAM system can persistently store ORAM-related data in NVM while solving the crash consistency problem without leaking more information.
Specifically, we first analyze the different components on the ORAM controller to determine the content that needs synchronous persistency and data consistency (details in Section \ref{s:background}).
Second, we analyze persistent atomic access and present different case studies that show what happens if data or other metadata is not persisted during a crash, and analyze the challenges of the problem and the system design goal.
Some feasible designs, in theory, are also presented. (details in Section \ref{sec:design_requirement}).
Third, we minimize the performance overhead due to the persistent write-back and propose an efficient and secure write-back scheme (details in Section \ref{Design}).
Finally, we test our scheme with the system without persistency and show in Section \ref{Evaluation}.

\section{Background}
\label{s:background}

In this section, we first describe the threat model. Second, we introduce the basics of ORAM and NVM.
{Then, we discuss the problems of traditional software-based persistence methods.}
Lastly, we describe how ORAM could be implemented on NVM based system.

\subsection{Threat Model}
We follow the conventional TCB boundary and assume that the system equips with a secure and tamper-resistance processor capable of computing without information leakage \cite{stefanov2013path,zhang2015fork,ren2013design,ren2015constants}. Everything on-chip is considered within the TCB boundary.
The off-chip main memory system is vulnerable to access pattern attacks, such as physically monitoring the visible signals on the printed circuit boards (including the motherboard and memory modules).
The address bus, the command bus, and the data bus are separate from  commodity DDR DIMMs in the system. As a result, the memory controller sends out the address and command in cleartext. Therefore, the attacker can still infer critical information even when the data bus is encrypted. By observing the access patterns such as access frequency, access type (read or write), and also the repeatability of accessing the same location, the attacker can obtain some leaked sensitive information in the program \cite{islam2012access}.

In some system settings, part of the main memory system can be considered as protected and free from most of security attacks. For example, with SGX \cite{johnson2016intel}, a small region in the memory called EPC can store pages safely. With cmov-based operation, the access to EPC region can be considered as oblivious too\cite{sasy2017zerotrace,ahmad2019obfuscuro}.
In this work, we discuss implementations under the two assumptions: 1) memory is fully untrusted; 2) memory has a partially trusted region. The different assumptions will change how ORAM metadata can be persisted without leaking information.
We discuss this issue in detail in Section \ref{sec:discussPosMap}.




\subsection{ORAM Basics} \label{Basic-Path-ORAM}
ORAM \cite{goldreich1987towards} is a security primitive that can hide the program's access pattern and accordingly eliminate information leakage. ORAM's basic idea is to access more blocks than the actual data we need, and shuffle the address space so that the access address becomes random.
With the ORAM controller in the secure processor, one memory access from the program is translated into an ORAM-protected sequence. ORAM protocol guarantees that any two ORAM access sequences are computationally indistinguishable. In other words, ORAM physical access pattern and the original logical access pattern are independent, which hides the actual data address with the ORAM obfuscation.
Since all ORAM access sequences are indistinguishable, an attacker cannot extract sensitive information through the access pattern.
Tree-based ORAM schemes, such as Path ORAM \cite{stefanov2013path} and Ring ORAM \cite{ren2015constants}, have improved the overall access and reshuffle efficiency greatly through cryptographic innovations. In this work, we focus on one of the most representative tree-based ORAMs, Path ORAM \cite{stefanov2013path}, which is the building block of many data oblivious frameworks, such as Obliviate \cite{ahmad2018obliviate}, Taostore \cite{sahin2016taostore} and Zerotrace \cite{sasy2017zerotrace}.

\begin{figure}[t]
  \centering
  \includegraphics[width=0.4\textwidth]{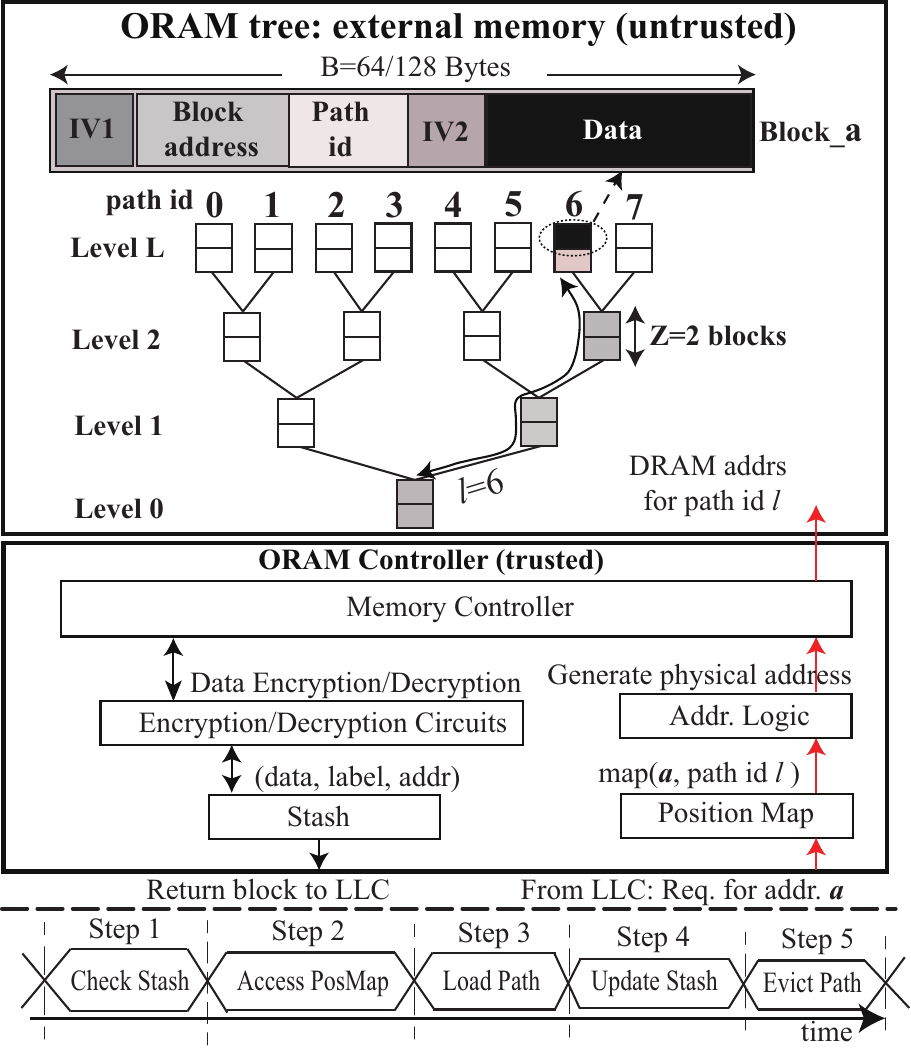}
  \vspace{-0.1in}
  \caption{Path ORAM construction and access protocol}\label{BasicOram02}
  \vspace{-0.1in}
\end{figure}

\subsubsection{Path ORAM Construction}
Logically, Path ORAM reorganizes the external memory into a binary tree (we refer to as the ORAM tree). Upon a memory request from the LLC, a full path of data blocks is fetched, as shown in Figure \ref{BasicOram02}.
The node in the ORAM tree is called a bucket and can hold $Z$ data blocks. The height of the ORAM tree is noted as $L$. In Figure \ref{BasicOram02}, we show an ORAM tree with 4 levels ($L=3$), and the bucket size equals to 2 ($Z=2$).
Each block inside the bucket contains the encrypted data content and a header that tracks the program address, path id, and initialization vectors(IV) used with AES counter mode encryption.
Dummy blocks are marked with a special program address $\perp$. Following \cite{fletcher2015low}, IV1 is used to encrypt the block's header, while IV2 is used to encrypt the data content.




On the trusted side, the ORAM controller converts the regular memory access pattern into ORAM sequences.
ORAM controller mainly includes a position map (PosMap), a stash, address translation logic, and encryption/decryption circuit.
The PosMap is a lookup table that stores the path id (leaf label) for a given logical address.
The stash is a small buffer that can hold a small number of data blocks \cite{ren2013design} during the path accesses.
The obliviousness of the access pattern is achieved by randomly remapping the path id of a data block after each access.


\subsubsection{Path ORAM Access Protocol} \label{Path-ORAM-Access-Protocol}
Next, we discuss the Path ORAM access protocol.
Given a memory request $a=(\emph{addr}, read/write, data)$ for data block $a$, the access steps of $ORAM(a)$ are as below:
\begin{enumerate}
    \item \textbf{Check Stash:} Check if the block $a$ is in the stash. If hit, fetch the data block to the processor if it is a read, or update the value if it is a write. If it is a miss, proceed to the next step.

    \item \textbf{Access PosMap:} The actual physical memory location of block $a$ is determined by checking the PosMap with $addr$, and a path id $l$ is returned.
    Then, randomly generate and update a new path id $l'$ for the accessed block $a$.

    \item \textbf{Load Path:} Load all blocks on path $l$ from the ORAM tree in the memory to the stash, decrypt them and find the block $a$.
    Then, return the block $a$ to the processor if it's a read operation, or update the value in the stash if it's a write operation.

    \item \textbf{Update Stash:}
    The path id of the block $a$ in the stash also needs to be updated to $l'$. In this case, data blocks in the stash have the most up-to-date value and path id.

    \item\textbf{Evict Path:}
    Evict data in the stash back to memory on path $l$. The basic rule of eviction is to fill as many blocks as possible that can be written to path $l$.
    If the real blocks are not enough, then pad with dummy blocks.

\end{enumerate}

\subsection{Persistent System with NVM}

Emerging NVM technologies, such as Phase-Change Memory (PCM), Spin-Transfer Torque (STT-RAM), and Memristor, are considered candidates for replacing conventional technologies such as DRAM and NAND Flash.
The Micron and Intel 3dXpoint-based Optane \cite{3dxpoint} has shown competitive performance, density and scalability with conventional technology.
When used as main memory, NVMs may provide {\em persistent memory}, where regular store instructions can be used to make persistent changes to data structures to keep them safe from crashes or failures. A great number of research efforts have sought to optimize recoverable or crash-consistent software (e.g.,
databases \cite{arulraj2015let,arulraj2017build}, file systems \cite{sehgal2015empirical,chen2013fsmac},
key-value stores \cite{yang2015nv,wu2016nvmcached}) for NVMs.

On the other hand, NVM systems still suffer from various security vulnerabilities. To provide data confidentiality, NVM can utilize lightweight encryption schemes \cite{young2015deuce,swami2016secret}; to detect and fix integrity issues, adopting Merkle tree and support its persistent updates have been recently studied \cite{zuo2019supermem,awad2019triad,yang2019no}. Access pattern leakage is another degree of vulnerability, and we can add obfuscation with the help of ORAM\cite{rakshit2018leo}.

\subsection{ORAM Systems with NVM}

While the main memory could be replaced with NVM, the on-chip cache and buffers are still using volatile memory for better performance and lower cost.
During a power failure, to ensure the on-chip contents can be flushed back to the NVM, Intel Asynchronous DRAM Refresh (ADR) scheme \cite{izraelevitz2019basic} provides the write queues are in the persistency domain.
However, when we have the ORAM controller sit between the write queue and the LLC, we need to consider how to persist the stash and PosMap, as they are not part of the persistency domain yet.

When the stash is volatile, after multiple ORAM accesses, a small number of data blocks in the stash become non-persistent. Such data blocks could contain the most up-to-date values for a given logical address. Consider that a failure happens during the execution, such content in the stash may be lost before they are written back to the NVM-based ORAM tree. The loss of data in stash not only causes a crash consistency problem but also causes the system to fail to correctly recover lost data blocks.
Similarly, the PosMap contains mapping information that determines where to locate a block in the main memory.
Each data block is given a path id, and it is not only associated with the block (in the header), but also stores in the PosMap.
As discussed in section \ref{Path-ORAM-Access-Protocol}, the updates on path id happen on multiple steps. If the PosMap is volatile, we will not be able to locate the block of interest in the main memory.

Furthermore, we identify that if the ORAM access needs to be recoverable, the data buffered in the stash and the PosMap needs to be persisted synchronously.
Otherwise, if the writebacks to NVM are asynchronous, it will cause data inconsistencies when we try to recover from a crash. We discuss the details of the writeback inconsistencies and design requirements in the next section.

\subsection{Problems with Software-based Crash-consistency Support}\label{SoftwareBasedInefficientandInsecurity}

{
Although traditional software-based mechanisms can be used to support crash consistency, we can know from \cite{ren2015thynvm} that their efficiency is very low.
}

{
The logging-based system \cite{volos2011mnemosyne, coburn2011nv}  maintains a backup copy of the original data in the log, and the log system redoes log (store new data) or undoes log (store old data).
Logging consumes much more NVM capacity than the original data, because each log entry is an original tuple of data and corresponding metadata (e.g., counter value, data address, etc.), and typically each memory record must be logged \cite{volos2011mnemosyne, coburn2011nv, ren2015thynvm}.
In addition, access to log recovery failure systems increases recovery time, reducing the advantage of a fast recovery system using NVM instead of slow block devices \cite{ren2015thynvm}.
Therefore, directly adopting logging-based technology to support the crash consistency of the ORAM system will not only bring more serious performance loss but also bring more memory overhead.
}

{
Similarly, a copy-on-write-based (CoW) system \cite{condit2009better, venkataraman2011consistent} always creates a new copy of the data to be updated.
The disadvantage of CoW is that the copy operation cost is expensive, and the stall time is longer \cite{seshadri2013rowclone}.
Since ORAM reads and writes multiple blocks along the path, if every accessed data block is to be copied, it will not only cause memory capacity overhead but also lead to more serious performance loss.
Also, additional NVM bandwidth is required due to the copy of redundant unmodified data blocks.  \cite{seshadri2015page, ren2015thynvm}.
There are dummy blocks in the ORAM tree, and it is a very serious performance loss if the accessed dummy blocks are backed up.
}

{In addition, software-based approaches may cause information leakage.
If the log is stored without protection, then the attacker will obtain the related access pattern or data information by peeking at the log, which will cause information leakage.}


\subsection{Design Challenges and Scope of This Work}

{To summarize, it is challenging to implement ORAM on NVM for three reasons: 1) ORAM is expensive in terms of memory access overhead; 2) simply replacing the  memory device to NVM cannot provide the ORAM accesses with crash consistency.
3) A simple software-based approach to support ORAM crash consistency has serious performance losses and security problems.}

In this work, we focus on enabling persistent ORAM system with low additional overhead. We believe that to achieve provable secure access pattern obfuscation, ORAM is required, and the cost of ORAM protocol can be further optimized with the cryptographic innovation.
On the other hand, ensuring crash consistency for the ORAM system is a critical problem to be solved by the computer architecture community when the memory system shifts to NVM technology.

\section{Design Requirements for Crash Recoverable ORAM}
\label{sec:design_requirement}
In this section, we discuss the design requirements for a recoverable persistent ORAM system. Simply replacing the main memory technology to NVM cannot guarantee consistent recovery. An ideal case would be that all on-chip buffers are built from NVM to write to the stash or position map is persistent immediately. However, as most of the on-chip components are still considered volatile, we identify a need to properly handle the volatile data in the ORAM controller to make the overall ORAM system persistent.

\subsection{Consistent Metadata Update}
\label{sec:requirement_consistency}
The ORAM accesses not only require updating the data block, but also the metadata associated with it, including the header and the position map entry.
Here, we define the \textit{consistent metadata update} requirement as follows:
when there is a crash happening at any ORAM access step, we can restart the ORAM access by identifying the target data block location in the NVM again. In other words, the path id information and other metadata should not be lost.

\begin{figure*}[t]
\centering
\subfigure
{
  \begin{minipage}{0.45\textwidth}
  \centering
  \includegraphics[width=0.95\textwidth]{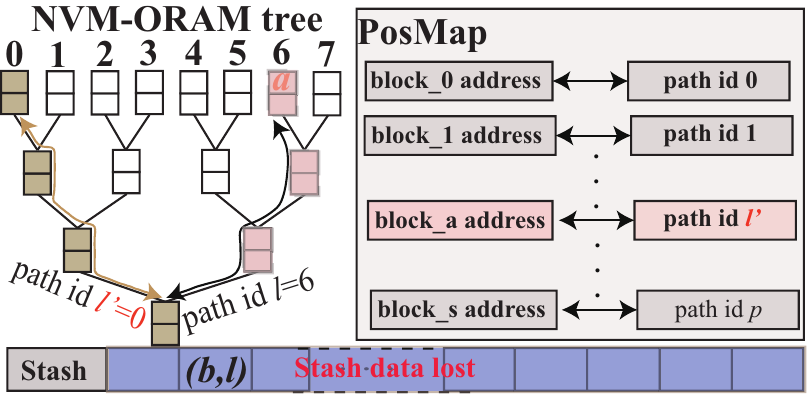}
  \end{minipage}
  \begin{minipage}{0.45\textwidth}
  \centering
  \includegraphics[width=1\textwidth]{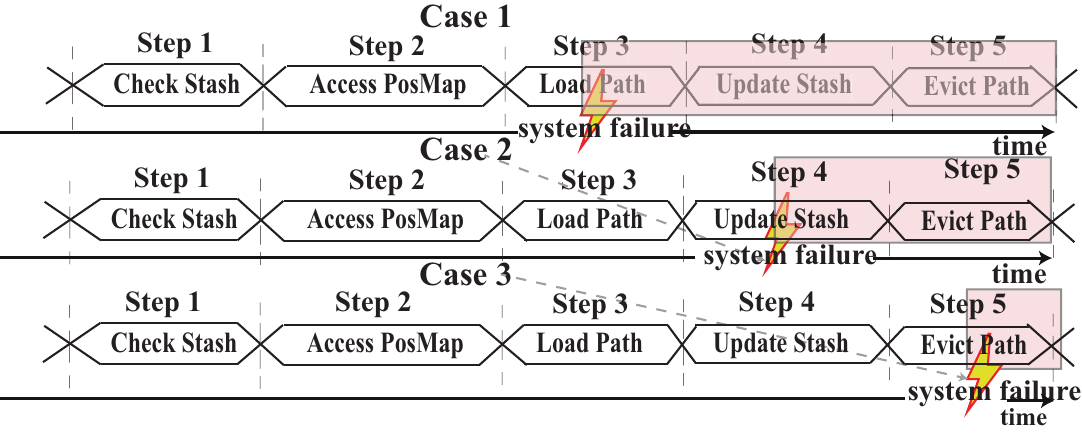}
  \label{fig:side:c}
  \end{minipage}
}
\vspace{-0.5em}
\caption{Step-by-Step diagram of NVM-based ORAM systems crash}
\vspace{-0.5em}

\label{exampleCrashORAM01}
\end{figure*}

Figure \ref{exampleCrashORAM01} demonstrates why consistent metadata update is desired.
In step 2 of an ORAM access,  a new path id is randomly generated for the target block, and the corresponding entry in the PosMap is updated.
If the metadata is not persisted consistently,
any crash happens after step 2 would possibly cause data inconsistency since the path id is changed.
We discuss the details by several case studies in Section \ref{sec:example}.

\subsection{Atomic ORAM Accesses to NVM} \label{Path-ORAM-Access-Atomicity}

Except for the consistent metadata updates, another design requirement for persistent ORAM is to preserve the access atomicity.
Here, we define the \textit{ORAM access atomicity} as follows:
The data in the stash and the metadata in the PosMap should reach persistency in an atomic way. If one of them is persisted while the other is not, the continued ORAM access is then out-of-sync.


The reason to have atomic ORAM access is that the metadata and the data correspond to the same actual memory request.
On a system failure, if only the content in the stash is persisted by writing back to the NVM-ORAM tree, the data content in the NVM-ORAM tree would be overwritten.
In this case, if the PosMap entries are not persisted yet, it is impossible to locate the new path id where the data is located.
A reverse example is if the metadata in PosMap is persisted, but the stash data is not, based on the new path id in PosMap, it is impossible to recover the lost data in the stash.  We also discuss the details of why atomicity is needed in Section \ref{sec:example}.


\subsection{Case Studies on Crash Recoverability}\label{sec:example}

To summarize, to ensure a recoverable ORAM access after a crash, we need to ensure the following requirements are met:

\begin{itemize}
 \item Ensure that the accessed data blocks in the NVM-ORAM tree are not lost during a crash. Data blocks in the stash that have not been evicted back into the NVM-ORAM tree can not be lost.
 \item The address and path id contained in each block evicted from stash to NVM-ORAM tree should be consistent with the metadata stored in the updated (persistent) PosMap, that is, consistent updates.
 \item The updated path ids of the accessed data in the PosMap, the data in the stash, should all reach the NVM atomically.
 Otherwise, there is a mismatch between the data persistency and metadata persistency.
\end{itemize}

Figure \ref{exampleCrashORAM01} shows an example that when the requirements are not met, during a crash, the NVM-based ORAM system could result in inconsistent status. We assume that $n$ ORAM accesses have been performed, so some data blocks remain in the stash, e.g., block $b$.
At the time of the crash, we are performing the $(n+1)$-th ORAM access. At step 2, the PosMap update is completed, i.e., the block $a$ is mapped to a new path id "$l'$" in PosMap. Then, on step 3-5, we could observe different types of inconsistencies due to the path id remapping process.


\vspace{+0.05in}

\noindent \textit{Case 1}: If the crash occurs in step 3 during the ORAM access, since the path id of block $a$ in PosMap has been updated ($l\rightarrow \textcolor{blue}{l'}$), and block $b$ has not been written back from the stash, no matter this metadata update is persisted or not, it violates the consistency and atomicity requirements.

\textit{a)} If the PosMap data has not been persisted, then the metadata in PosMap is restored to the last persistent state.
In the worst case, if the metadata in PosMap is not persisted after performing $n$ ORAM accesses, it returns to the initial state when the NVM-based ORAM system starts. The data blocks distribution in the NVM-ORAM tree has already changed after $n$ ORAM accesses.
If the program continues to perform ORAM access based on the old metadata in the PosMap, it will cause access errors.

\textit{b)} If the metadata in PosMap has been persisted after the update, then we can always retrieve the most up-to-date metadata when a crash happens. In this case, we retrieve path $\textcolor{blue}{l'}$ for block $a$.  However, $a$ is never written back to path $\textcolor{blue}{l'}$. Therefore, even with the persisted new path id, we cannot fetch the data from NVM again.


\textit{c)} Regardless of whether the metadata in PosMap has been persisted, the data blocks stored in the volatile stash are lost, including the dirty ones with new values.

\vspace{+0.05in}
\noindent \textit{Case 2}: If the system crash occurs in step 4, the good news is that block $a$ is already fetched and path $l$ has been fetched into the stash so that this particular access may succeed. However, similar to the {case 1}, the content in the stash would be all lost.

\vspace{+0.05in}

\noindent \textit{Case 3}: If the system crash occurs in step 5 of ORAM access, or before the next ORAM access, it may cause inconsistent data updates. Step 5 is to write data back to the NVM-ORAM tree, and the operation is a natural data persistency operation. We discuss the following scenarios that may happen:

\textit{a)} The stash content is lost, similar to case 1 and 2.

\textit{b)} During the path eviction process, it is possible that some data blocks have been written back to the NVM-ORAM tree while some are not. This will cause non-atomic data write-backs to the ORAM tree and overwrite some of the real data blocks.

 \begin{figure}[t]
  \centering
  \includegraphics[width=0.45\textwidth]{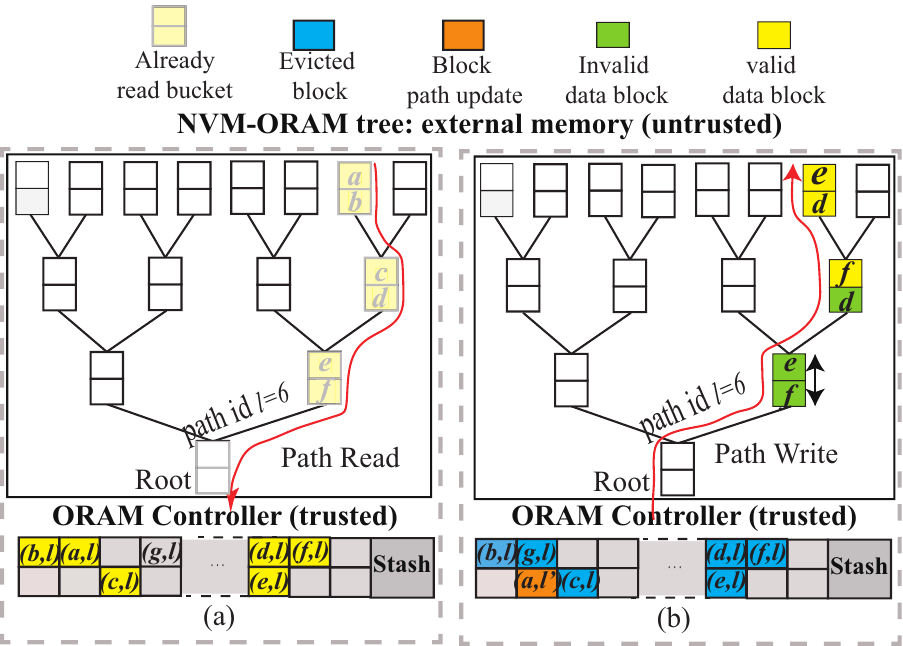}
  \caption{Data overwritten by partially path writeback}\label{TraditionalORAMWriteBack01}
\end{figure}

Figure \ref{TraditionalORAMWriteBack01}(a) shows the data fetched from the NVM-ORAM tree into the stash.
In the original write back process, the target block $a$ in the NVM-ORAM tree is overwritten by the write back process, and the position of other blocks also changes, as shown in Figure \ref{TraditionalORAMWriteBack01}(b).
Obviously, if the system crashes after the completion of the writeback and before the next ORAM access, block $a$ will be lost and unrecoverable.
If a system crash occurs during write back, more data blocks may be lost.
If the system crashes when writing back data block $g$, the data in the stash is lost, and the data blocks $a$, $b$, and $c$ on the NVM-ORAM tree have been overwritten by blocks $e$, $d$, and $f$, respectively.
Data blocks $a$, $b$, and $c$ are lost and cannot be recovered, as shown in Figure \ref{TraditionalORAMWriteBack01}(a).

Through the in-depth case study, we understand the design requirements for a recoverable persistent ORAM system. We do not want the stash content to be lost; meanwhile, we would like the updates on PosMap to be consistent with the contents in the stash; further, we would like to ensure the atomicity of data and metadata writebacks.


\subsection{Design Challenges}
\label{sec: challenges}

To this end, we have identified the design requirements for ORAM with crash consistency built on NVM. Meeting the requirements above can ensure a consistent data recovery when the system experiences a crash. In addition, the data persistency process should come with a {low performance} overhead and do not introduce significantly more writebacks to the NVM system.  This section describes the design challenges when we would like to fulfill the design requirements above.

\noindent \textit{Challenge 1: } How to ensure that the data blocks fetched from the NVM-ORAM tree to stash are not lost during the system crash when the on-chip buffers are volatile?
%


As we discussed ahead, before eviction, the NVM-ORAM tree still contains persistent data for the current ORAM access. However, we cannot re-read the block of interest back to the stash after the crash because the path id has been changed after each data block is being accessed. A simple solution is to back up the data block back through writing it to its original path. Clearly, in this way, the security of ORAM will be wrecked. If we can utilize the natural persistent eviction operation to store backup data in the original path and maintain the random ORAM path id remapping, we can ensure both crash consistency and security.

\noindent \textit{Challenge 2: } How to control the data and metadata are persisted atomically?

For each ORAM access, the data in stash and metadata in PosMap needs to enter the persistence domain at the same time to ensure the access atomicity. However, merely using one ADR-based write queue without proper control will not guarantee that the persistent writebacks follow the ORAM sequences. As a result, we will need to revisit the control logic that can enforce the atomic writebacks.

\noindent \textit{Challenge 3: } How to incorporate atomic ORAM access into the original ORAM protocol?

The atomic ORAM access with persistency requires additional writebacks to the main memory, in addition to the original one-step eviction. As a result, the access pattern to the memory system is slightly changed. Especially for persisting the PosMap, directly write the updated entry back to NVM may still expose which data has been accessed, if the PosMap in the NVM is not protected.

\section{EHAP-ORAM: The Design of Crash Consistency ORAM} \label{Design}
In this section, we present EHAP-ORAM, a new crash-consistent architecture designed for persistent ORAM.
The EHAP-ORAM includes a new ORAM controller architecture that requires the necessary hardware support to protect the accessed data blocks from loss and consistency with metadata updates in PosMap.
Further, we incorporate the persistent atomic writebacks into the ORAM protocol and analyze how the crash consistency can be achieved. The hardware and protocol innovations ensure that the persistency is done correctly and do not destroy the ORAM obfuscation capability.


\subsection{Architecture Overview}
To protect the crash consistency of the NVM-based ORAM systems, we design the EHAP-ORAM controller from four aspects.
\begin{enumerate}
\item It does not destroy the access sequence of the original Path ORAM access.
\item The EHAP-ORAM controller does not significantly impact the performance of ORAM access.
\item It can maintain the consistency of the accessed data blocks and the metadata updates in PosMap.
\item If the system crashes, when restarting or restoring the system, the lost data blocks and related metadata can be recovered effectively and correctly.
\end{enumerate}

\begin{figure}[t]
  \centering
  \includegraphics[width=0.45\textwidth]{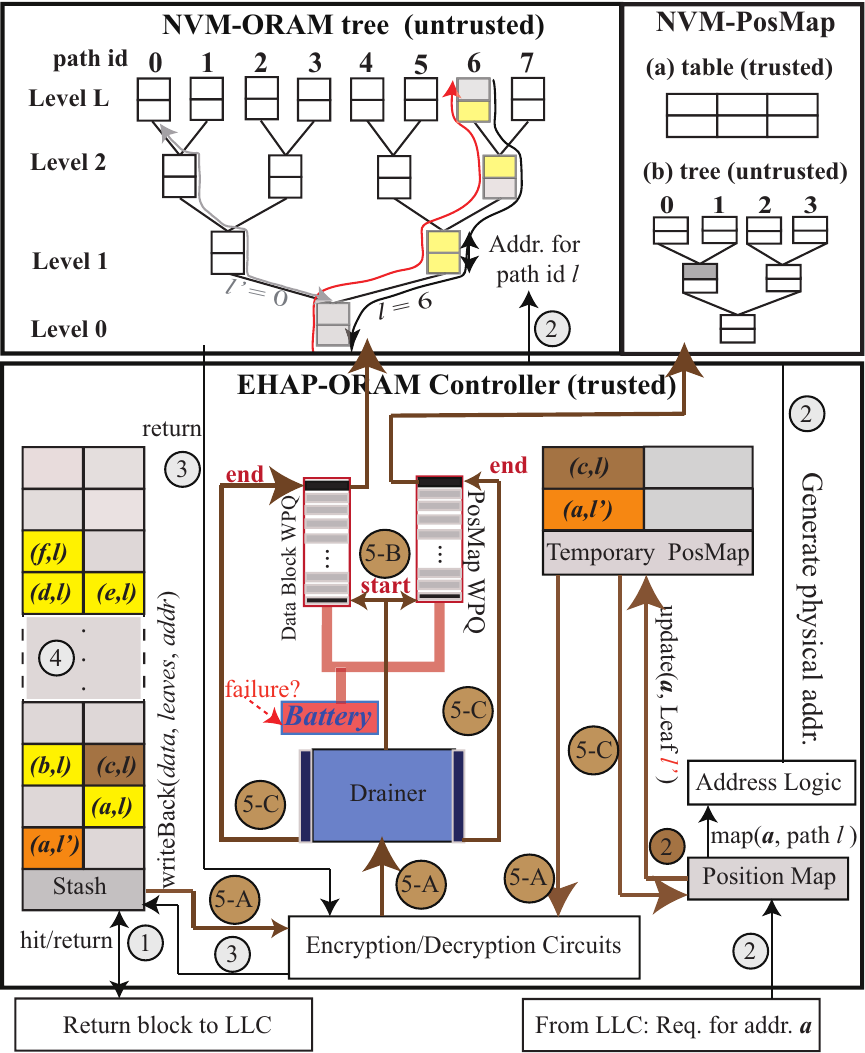}
  \caption{EHAP-ORAM system architecture.}\label{oram01}
\end{figure}

We show the proposed EHAP-ORAM architecture in Figure \ref{oram01}, {similar to \cite{liu2018crash} and \cite{yang2019no}}.
Besides the basic components of the existing ORAM controller (i.e, a stash and a PosMap), four components are added: a \textit{Drainer}, a \textit{Temporary PosMap} and two \textit{write pending queues(WPQ)}.
The basic functions of the newly added components are explained as follows:


The drainer is connected to the encryption/decrypting circuit and dispenses the data evicted from the stash and temporary PosMap to the two corresponding WPQs.
And drainer is responsible for issue control "start" and "end" signals to control the queue receiving data and the signals persisted to the NVM.

The temporary PosMap stores the reassigned path ids of the accessed target data blocks.
Specifically, according to the access protocol of Path ORAM, we know that each time the ORAM controller touches a target data block, a new path id is assigned to it (see {step 2} of \ref{Path-ORAM-Access-Protocol}).
At this time, the address of the accessed target blocks and the corresponding new path id will be stored in the temporary PosMap to wait for the data to be persisted. As long as the temporary PosMap is not merged with the main PosMap, we do not overwrite the original path id that has been persisted already.

The two WPQs are considered as persistence domain, similar to the ADR-based approach.
They are both backed with battery, therefore, during a power failure, the contents can still be flushed back to the NVM atomically.
The data blocks WPQ is used for storing evicted data blocks from stash.
Each time when the data blocks are evicted from the stash, they then enter the data blocks WPQ to achieve atomic persistence.
The PosMap WPQ is used to persist recently changed path ids corresponding to the data blocks evicted from the stash. The content in PosMap WPQ comes from the temporary PosMap.
{Note that the backup battery does not need to be embedded on-chip, it only needs to support the two WPQs on-chip through the wire connection.}

The size of temporary PosMap and WQPs depends on the ORAM configurations.
First, the temporary PosMap size needs to match the stash size to avoid overflow. Consider the stash with a size of $C$, it may store up to $C$ dirty data blocks that have been accessed recently. Therefore, the temporary PosMap should also be able to store $C$ entries. In our experiments, we follow \cite{ren2013design} and set the $C=200$. In this case, the additional storage overhead of the temporary PosMap is $C$ address to path id mapping entries.
Second, the size of the data block WPQ depends on the height of the ORAM tree.
The maximum number of write-back data blocks from the stash each time equals to the total number of data blocks on a path.
Therefore, the minimum size of the data block WPQ is equal to $Z*(L+1)$ blocks.
Finally, the size of PosMap WPQ depends on the number of blocks in data block WPQ that have a changed path id.
Considering the worst case, that is, all blocks in the data block WPQ have a new path id, then the maximum capacity of the PosMap WPQ is $Z*(L+1)$ path ids. On an average case, only a small number of real data blocks in the stash are fetched by demand.
Therefore, the PosMap WPQ could be a lot more smaller.



The EHAP-ORAM controller adds the needed hardware to support persistent ORAM accesses.
Next, we discuss how controllers work with the ORAM protocol and how to achieve crash consistency.

\subsection{EHAP-ORAM Workflow}

We then describe the EHAP-ORAM workflow in this section.
To provide the ORAM system with crash consistency, we revisit the basic Path ORAM workflow and carefully integrate the persistent operations during the ORAM access.
The updated ORAM access protocol still follows the main workflow without leaking information. The circled numbers in Figure \ref{oram01} represent the dataflow corresponding to each step of the EHAP-ORAM protocol.

\subsubsection{EHAP-ORAM access protocol} \label{PS-ORAM Access}
Given a memory request $a=(\emph{addr}, read/write, data)$ for data block $a$, the five access steps of \textit{EHAP-ORAM(a)} are as below:

\begin{enumerate}
    \item{\textbf{Check Stash:} This step remains unchanged as Step 1 in Section \ref{Basic-Path-ORAM}.} If the block $a$ is not in the stash, proceed to the next step.

    \item{\textbf{Access PosMap and Backup Label:}
    Similarly, we check PosMap with $addr$, and $l$ is returned as the target path id.
    Then, the data block $a$ is remapped to the new path id label $l'$. Instead of overwriting the $(a,l')$ directly in the PosMap, EHAP-ORAM stores the new path id $l'$ in the \textit{temporary PosMap}.
}
    \item{\textbf{Load Path:} This step is unchanged from the original Step 3 in Section \ref{Basic-Path-ORAM}.
    }

      \item{\textbf{Update Stash and Backup Data:} Since the new label $l'$ has been reassigned to the target data block $a$ in step 2, the path id in the header of the data block $a$ fetched from the NVM-ORAM tree to stash is now updated to $l'$.
      Meanwhile, the original data block $(a,l)$ is copied in the stash as a backup block (similar to the concept of shadow block in \cite{zhang2018shadow}).
      In this case, we can make sure the backup block will be written back to path $l$ during the eviction. Later on, when block $a$ is persisted on path $l'$, this backup data block can be marked as invalid. }

    \item{\textbf{EHAP-ORAM Eviction:} Lastly, the eviction needs to be done properly to ensure crash consistency. We describe the details of the persistent evict path operation in Section \ref{Eviction write back}.}
\end{enumerate}

Note that the main function of the backup data block generated in Step 4 of EHAP-ORAM access is to recover the data block lost after the crashed system.
We analyze how to recover lost data in Section \ref{Data recovery}.

\subsubsection{EHAP-ORAM eviction in detail}\label{Eviction write back}
The EHAP-ORAM eviction is the main step of writing the data or metadata from volatile on-chip components back to the persistent NVM system. Due to the several design requirements we would like to achieve, the new eviction operation also contains several sub-steps that closely interact with the new components in the EHAP-ORAM controller.


\begin{itemize}
\item{\textbf{Step 5-A}
The data blocks that need to be written back from the stash are identified first. Because EHAP-ORAM loads the path $l$ in Step 3, the eviction path is also $l$.
In Figure \ref{oram01}, the yellow and brown blocks are identified and they will be encrypted. Note that the backup block $(a,l)$ is also included as an eviction candidate.
Meanwhile, if the data block's path id has been changed, the corresponding metadata entries in the temporary PosMap are identified. In this example, the entry $(c,l)$ is identified and will be encrypted. The block $c$ was previously fetched and path $l$ is its new path id.
}
\item{\textbf{Step 5-B} Once the eviction data blocks and metadata are ready from encryption, the drainer sends the ``start" signal, and the candidate data blocks and PosMap entries are loaded into the two corresponding WPQs.
Note that the ``start" signal controls both WPQs, as such, the data and metadata can be load into the persistence domain atomically.
    }
\item{\textbf{Step 5-C}
When the data and metadata for this eviction round are all in the WPQ, an ``end" signal is sent to both WPQs, meaning that the ORAM eviction is now atomic. Then the two WPQs are flushed back to the NVM-ORAM tree and the PosMap in the NVM. Note that the storage format of PosMap depend on the threat model: if the PosMap is kept in a trusted region in the NVM, then the write back can be done through direct updates to the table; if the PosMap is not kept in a trusted NVM region, recursive PosMap is needed to keep the writebacks secure. Figure \ref{oram01} shows the two formats of storing PosMap in memory securely. We discuss the options to implement the two PosMap WPQ flushing cases in Section \ref{sec:discussPosMap}.

}
\end{itemize}

Tracking the dirty PosMap entries and only putting them into the WPQ can greatly reduce the performance overhead. Otherwise, for all $Z*(L+1)$ blocks on the path, we need to flush $Z*(L+1)$ PosMap entries as well (refers to full writeback in our experiments). With our on-demand writeback scheme, we achieve the same level of persistency while removing the majority of data writes (see section \ref{Evaluation:Methodology}).



\subsection{Data Recovery Consistency Analysis} \label{Data recovery}
In this section, we show how EHAP-ORAM can guarantee a consistent crash recovery through case studies. We revisit the three cases in Section \ref{Basic-Path-ORAM} and analyze why the prior issues are addressed.

\vspace{+0.05in}
\noindent \textit{Case 1}: In the original Path ORAM, PosMap has been updated before step 3 of each ORAM access.
As a result, when the system crashes during step 3, data blocks stored in the volatile stash are all lost.
Therefore, it will cause a crash consistency problem because the data in the volatile stash is not persisted in time.

With EHAP-ORAM architecture, since step 2 is enhanced, the new path ids of the accessed data blocks are not committed directly into the PosMap but into the temporary PosMap (volatile).
Therefore, if the EHAP-ORAM system crashes in step 3, the data in the temporary PosMap, and stash will all be lost at the same time.
During the recovery process, the ORAM controller can re-read this path id before remapping again with consistent path id in the PosMap.
Therefore, when performing this ORAM access again, the matching PosMap can still correctly access the data of interest in the original path from the NVM-ORAM tree.


\vspace{+0.05in}

\noindent \textit{Case 2}: When the system crash occurs at step 4 of the ORAM access, the scenario is similar to case 1.
The difference is that the ORAM controller has fetched data blocks from a path to stash, so the data blocks on that path are marked as invalid. Invalidate data blocks in the NVM-ORAM tree only happen with some updates on metadata, not the actual data content, therefore, there is no data loss or mismatch happening.
During the recovery, the ORAM controller only needs to restore the data that has been marked as invalid to a valid during the read path. Then, the lost data can be recovered from the data content region.

\vspace{+0.05in}

\noindent \textit{Case 3}: If the ORAM system crash occurs in step 5 of the ORAM access or before the next ORAM access, as discussed before, it may cause inconsistency with partial writebacks (either data or metadata). As a result, some valid data along the path are no longer recoverable. Also, lost data in stash and PosMap scenario is similar as Case 1 and 2.

We create the backup block for accessed block and write it back to the original path $l$ together with other data blocks to solve the overwritten problem.
At the same time of writing back, PosMap does not update the path id of the target block that has not been evicted from stash, so the target block's original path id is still stored in PosMap (Section \ref{PS-ORAM Access} Step 3).
If the system crashes at this time, the target blocks that have not been evicted in the stash are lost, but their backup blocks can still be found and restored in the NVM-ORAM tree.

In addition, the added on-chip WPQs can ensure the volatile data in stash and PosMap enter the persistence domain at the same time. We do not need to worry about the content in the stash,and PosMap is gone with a crash.

If the system crashes before the ``end" signal is received by the write pending queue, the original data blocks on the write-back path still exist and will not be overwritten, so the data can be recovered.
Therefore, with EHAP-ORAM writeback operation, the data blocks in stash and PosMap can be consistent, and the data blocks lost after the system crash can be effectively recovered.

\subsection{Ways to Store and Persist PosMap in Memory} \label{sec:discussPosMap}
PosMap is the key component in ORAM system, as it stores all mapping information for each memory request. Phantom \cite{maas2013phantom} is the first hardware ORAM prototype built on FPGA. Since the FPGA memory is relatively small, the Phantom design stores the entire PosMap on the chip.
As the capacity of the ORAM tree increases, Phantom needs to use multiple FPGAs to store the metadata in PosMap. In either cases, the PosMap is always on-chip, therefore, storing the contents in plaintext is acceptable.

However, if the ORAM tree size is large, it is hard to store the entire PosMap on-chip. For example, a $4$GB ORAM tree with $128$bytes and $Z=4$ requires a $93$MB PosMap size \cite{ren2013design}.
In order to solve the problem of large PosMap size, recursive ORAM is proposed \cite{ren2014unified,fletcher2015freecursive}. In this way, the PosMap in untrusted main memory is also stored as a small ORAM tree, while the on-chip PosMap is a cache for most recently used PosMap entries. Update the PosMap in the memory requires a small ORAM tree write path operation.

A more ideal case would be, the PosMap can be stored in a trusted memory region and any read or write operations to the PosMap are free from most of security vulnerabilities \cite{sasy2017zerotrace,ahmad2018obliviate,ahmad2019obfuscuro}. In this case, a cmov-based oblivious update is desired to further obfuscate the access pattern to the PosMap. The oblivious PosMap update generates fake addresses for all entries in the PosMap, but only the updated entries will be actually written.

In this work, we consider both cases of implementing and accessing PosMap on NVM main memory. We implement the recursive ORAM and PosMap accesses following \cite{fletcher2015freecursive}.
Also, we consider the PosMap is kept at a on-chip secure region (similar to \cite{maas2013phantom}) and cmov-based PosMap updates can ensure the writebacks are still oblivious.




\subsection{Security Analysis}
ORAM is designed to hide the original program's memory access pattern, and its security depends on the independence of the label sequence, randomness, and the same length of the access sequence \cite{stefanov2013path}.
In EHAP-ORAM, we modify the step 2,4 and 5 of ORAM access for the add-on persistency. However, we do not modify the random remapping process and the redundant sequences of ORAM access. The added components and data block backup steps all happen on the trusted ORAM controller side. Therefore, the modifications do not leak any access pattern information, or cause stash/ORAM tree capacity overflow.

\noindent \textit{Claim 1: Step 2 does not leak additional information.} The backup label operation happens inside of the ORAM controller, which is inside of the trusted boundary.

\noindent \textit{Claim 2: Step 4 does not leak additional information or cause overflow.} The backup data block is written back to the original path each time. Therefore, the stash occupancy does not change after each ORAM access.
When the block is written back to its new path, the previous copied block is marked as invalid, so occupied memory space is freed again. As a result, we do not increase the stash and ORAM tree overflow probability. A similar use case has been discussed in \cite{zhang2018shadow}.

\noindent \textit{Claim 3: Step 5 does not leak information during the writebacks.}
The data blocks written back from WPQ remain the same as the baseline Path ORAM.
As for the security of PosMap, in this work, we consider two situations to protect PosMap.
When the PosMap is stored in an SGX-like trusted memory region \cite{johnson2016intel}, the CMOV-based PosMap update approach \cite{ahmad2019obfuscuro,sasy2017zerotrace} is adopted to ensure the obliviousness. On the memory address bus, all entries in the PosMap is touched, but only the ones that require changes are written with new values.
If no trusted memory region is available, we store the PosMap recursively \cite{fletcher2015freecursive}, and the writing back one path id updates involves a small PosMap ORAM path write.
Hence, EHAP-ORAM PosMap writeback does not introduce additional access pattern leakage.

Summarize, EHAP-ORAM architecture and its access protocol support crash consistency without leaking additional information on access patterns.

\section{Evaluation} \label{Evaluation}

In this section, we first describe the relevant settings for experimental evaluation.
Then, the designs of the experimental evaluation are described.
Finally, the detailed evaluation results of each experimental design are given.

\subsection{Methodology}
\label{Evaluation:Methodology}
For the simulation of hardware design, we used a cycle-accurate Gem5 simulator \cite{binkert2011gem5} for modeling and simulation.
The detailed memory access modeling of the NVM-ORAM tree uses NVMain 2.0 \cite{poremba2015nvmain}.
The simulation system consists of a x86-64 {in-order processor} running at 3.2GHz.
Without losing generality, we use phase-change memory (PCM) with 4GB capacity \cite{choi201220nm, poremba2015nvmain}.
{Table \ref{table001}} summarizes the list of processor, ORAM controller, and main memory configurations.
For other system-related parameters, we use the default values of gem5 and NVMain 2.0.

The memory capacity of the basic NVM-ORAM tree is 4GB \cite{wang2018d}.
We modify the controller interface of NVMain 2.0, simulate and test the EHAP-ORAM system proposed in this work, and compare it with basic ORAM technology \cite{stefanov2013path}.

To be able to accurately evaluate the performance of all aspects of the EHAP-ORAM system, in the experimental test of this work, the classic SPEC 2006 \cite{henning2006spec} benchmark suite was selected as the workloads.
We selected 14 workloads from SPEC 2006 to test the EHAP-ORAM system.
The applications are described as Table \ref{MKPI}.
We use a method similar to that in \cite{komalan2018main} to consider 5,000,000 samples per trace in all selected SPEC simulation programs.
Note that in the experiment of this work, we will not simulate the performance impact of data encryption.

\begin{table}[t]
\caption{Experimental Setting Configurations}
\centering
\footnotesize
\subtable[\textbf{Core, on-chip cache}]{
 \begin{tabular}{|l|l|l|}
\hline
   Core type & in-order \\
   \hline
   Core number  &  $1$ Core, $1$ Thread\\
   \hline
   Core frequency  & $3.2$ GHz \\
   \hline
   L1 I/D cache & $32$KB/$32$KB, $2$-way \\
   \hline
   L2 cache & $1$MB shared, $8$-way\\
\hline
\end{tabular}
\label{firsttable}
}
\qquad

\subtable[\textbf{ORAM controller}]{
\begin{tabular}{|l|l|l|}
\hline
   Controller clock frequency & $3.2$ GHz \\
   \hline
   Data block size  &  $64$B \\
   \hline
   Data ORAM capacity  & $4$GB ($L=23$) \\
   \hline
   Block slots per bucket ($Z$) & $4$ \\
   \hline
   Stash size ($C$) \cite{ren2013design}& $200$ (blocks)\\
   \hline
   Temporary PosMap size ($C_{tPos}$) & $200$ (path ids)\\
\hline
\end{tabular}
\label{secondtable}
}
\subtable[\textbf{Main Memory (NVM)}]{

\begin{tabular}{|l|l|l|}
\hline
   Memory type & $4$GB PCM, $400$ MHz \cite{choi201220nm},                                    \\
                &$t_{RCD}/t_{WP}/t_{CWD}/t_{WTR}/t_{RP}/t_{CCD}$\\
                &=48/60/4/3/1/2  \\
\hline
\end{tabular}
\label{Thirdtable}
}
\label{table001}
\end{table}

\begin{table}[t]
\caption{Workloads and their MPKI (Memory Accesses Per 1000 Instructions)}
\centering
\footnotesize
\setlength{\tabcolsep}{3mm}{
\begin{tabular}{|l|l|l|l|}
\hline
Workload & MPKI &Workload & MPKI\\
\hline
\hline
   401.bzip2 & 61.16   &464.h264ref   & 19.74  \\
   \hline
   403.gcc &  1.19   &471.omnetpp   &  7.84\\
   \hline
   429.mcf& 4.66     &483.xalancbmk   &  8.99\\
   \hline
   445.gobmk & 29.60   &444.namd   &  8.08\\
   \hline
   456.hmmer & 4.53    &453.povray   &  6.12\\
   \hline
   458.sjeng & 110.99  &470.lbm   &  18.38\\
   \hline
   462.libquantum & 18.27   &482.sphinx3   &  17.51\\
\hline
\end{tabular}
\label{MKPI}
}
\end{table}

Based on the hardware architecture of the persistent ORAM controller, four different persistent ORAM system protocols are tested and compared with the basic non-recursive/recursive ORAM system protocols.
\begin{itemize}
\item{
\textbf{No Persistency (Baseline):} We apply the basic ORAM protocol to an ORAM system with NVM, which has no application persistency design, i.e., Baseline.
}

\item{
\textbf{On-chip with NVM (FullNVM):}
 Although it is theoretically feasible to replace NVM with on-chip storage, it is impractical to manufacture according to the existing process technology.
But, to verify the correctness of our conjecture, we still do the experiment and simulation on this design in gem5.
In this design, we persist the updated PosMap metadata to the NVM-PosMap in the trusted region.
}

\item{
\textbf{Full Persistency (FP)}:
In this design, the metadata in PosMap is stored on the off-chip untrusted NVM.
To reduce the leakage of information and make the data persisted on the NVM-PosMap more secure, we build the NVM-PosMap into a tree structure.
We persist the metadata in the updated PosMap to the NVM-ORAM tree in a method similar to \cite{le2020tale}.
Each time the ORAM eviction write-back is executed, a path on the corresponding NVM-PosMap tree will be written with metadata.
Different from recursive ORAM design, except that the system initialization requires reading metadata from NVM-Posmap into the on-chip PosMap, the system only performs write operations when performing ORAM access.
}

\item{
\textbf{On-demand Persistency (EHAP-ORAM)}:
In FP design, much redundant metadata will be written back to NVM PosMap to enhance security.
Actually, when updated PosMap metadata is persisted to the NVM-PosMap in the trusted region, redundant data writes can be reduced.
Specifically, the EHAP-ORAM system performance can be improved by persisting only with the changed data.
So, in this design, the metadata in PosMap is stored in a trusted NVM region off-chip.
}

\item{
\textbf{Recursive ORAM without Persistency (Rcr-Baseline):} In the same way as the NPr design, we apply the basic recursive ORAM protocol to an ORAM system with NVM.
Recursive ORAM \cite{fletcher2015freecursive} without a persistency design serves as the baseline for persistent recursive ORAM systems.
}

\item{
\textbf{Recursive ORAM with On-demand Persistency (Rcr-EHAP-ORAM):}
Recursive ORAM can effectively solve the problem of the size of PosMap, we also add a persistency design to recursive ORAM and conduct simulation tests on it.
Following \cite{ren2013design, zhang2015fork}, a $4$GB ORAM with a block size of $64$ bytes and $Z = 4$ has a PosMap of 192 MB.
Therefore, when we build the PosMap NVM-ORAM tree, we need a memory size of $192*2=384$ MB.
The block size of the PosMap ORAM tree is $64$ bytes and each bucket has $Z=4$ blocks, the number of tree levels is $21$.
}
\end{itemize}
\subsection{Evaluation Results} \label{EvaluationResults}
In this subsection, we evaluate the performance impact of different designs (see section \ref{Evaluation:Methodology} for details).
In different channel systems, we use the basic ORAM protocol without non-persistent (NPr) design as the baseline.
In this experiment, we set the data block and PosMap WPQ sizes to $24*4=96$ (blocks), {i.e., $96 \cdot 64$ bytes $= 6144$ bytes} and $24*4=96$ (path ids), {i.e., ($96 \cdot (32+24))= 5376$ bits $= 672$ bytes}, respectively.
To simplify the discussion, we focus on compare and analyze  different workloads under different system configuration parameters.
\\
\begin{figure}[b]
\subfigure[Performance comparison of different designs in single-channel system.]
{
  \begin{minipage}{0.5\textwidth}
  \centering
  \includegraphics[width=0.95\textwidth]{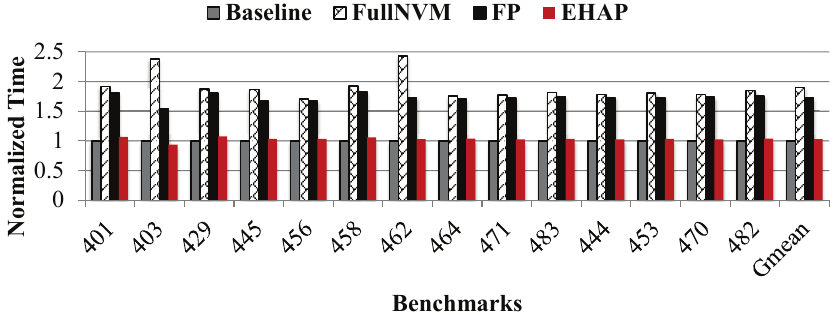}
  \label{1channel-NVMstash-RecursiveORAM-NVMORAM}
  \end{minipage}
}
\subfigure[Performance comparison of EHAP-Recursive ORAM in single-channel system.]
{
  \begin{minipage}{0.5\textwidth}
  \centering
  \includegraphics[width=0.95\textwidth]{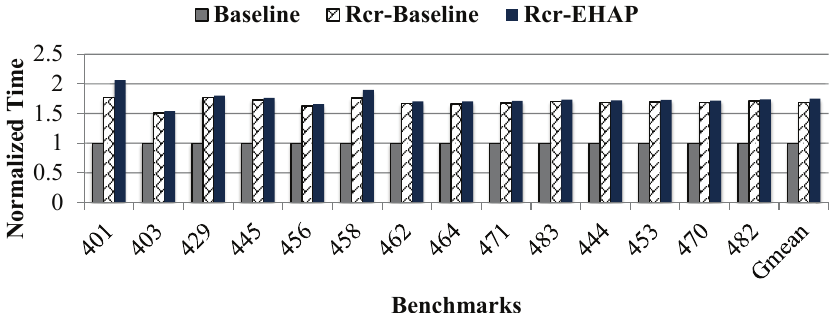}
  \label{NVMRecursiveORAMPerformance}
  \end{minipage}}
\vspace{-0.8em}
\caption{Performance comparison ($Z=4$, $channel = 1$, $core=1$).}
\label{PerformancePsORAMandRecursiveORAM}
\end{figure}

\begin{figure*}[t]
\subfigure[Compare the number of reads]
{
  \begin{minipage}{0.5\textwidth}
  \centering
  \includegraphics[width=0.95\textwidth]{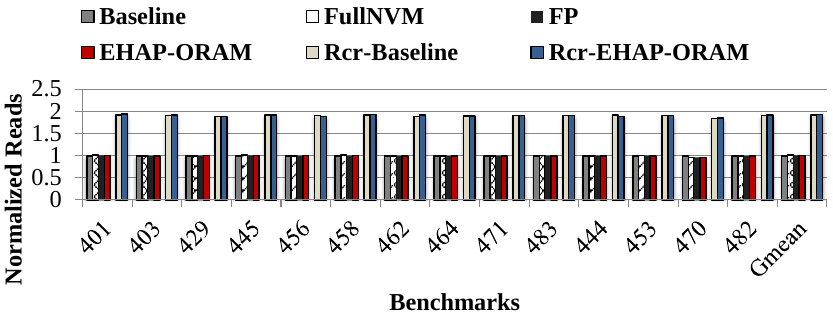}
  \label{NumberofNormalizedReads}
  \end{minipage}
}
\subfigure[Compare the number of writes]
{
  \begin{minipage}{0.5\textwidth}
  \centering
  \includegraphics[width=0.95\textwidth]{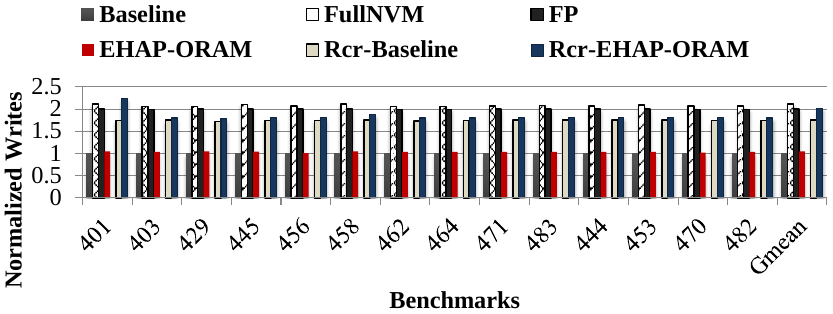}
  \label{NumberofNormalizedWrites}
  \end{minipage}}
\vspace{-0.8em}
\caption{Comparison of reads and writes of different designs.}
\label{1channel-NVMstash-RecursiveORAM-NVMORAM-AccessNumber}
\end{figure*}

\subsection{System Performance}
\textbf{A. Single-Channel Performance.}
Figure \ref{PerformancePsORAMandRecursiveORAM} shows the standardized access latency under different workloads for basic ORAM and recursive ORAM with persistency designs in a single channel system.

Figure \ref{1channel-NVMstash-RecursiveORAM-NVMORAM} illustrates the impact of different designs on the performance of basic ORAM systems when performing different workloads.
We have the following observations:
(1) FullNVM, compared with traditional ORAM systems and other designs, the performance cost of the FullNVM design is the largest, with an average performance drop of approximately 90.54\% {(i.e., a $1.9$\textbf{x} execution time)}.
The storage modules in the ORAM system are all replaced by NVM, every read and write is performed on NVM, which is currently relatively slow than traditional volatile storage, the performance loss is naturally high.
(2) FP, the performance of the full persistency design is slightly better than that of FullNVM design.
Compared with the baseline, the average performance is reduced by 73.18\%, performance improved by 17.36\% over FullNVM.
This is mainly because the traditional non-volatile stash on the chip side is faster than NVM to read/write.
However, since the persistency needs to be performed every time an ORAM access is performed, its performance is worse than the EHAP-ORAM design.
(3) EHAP-ORAM, compared with the baseline, the performance loss of EHAP-ORAM design is only slightly reduced, about 3.36\%.
Compared with FullNVM and FP, the performance of EHAP-ORAM is improved by 87.18\% and 69.82\%, respectively.
This is because the EHAP-ORAM design only persists metadata related to the target block data that has been updated, reducing unnecessary redundant data persistency.

Figure \ref{NVMRecursiveORAMPerformance} shows the performance loss of persistent recursive ORAM compared to basic ORAM and non-persistent recursive ORAM.
Obviously, both the basic recursive ORAM (Rcr-Baseline ) and Rcr-EHAP-ORAM have a high-performance overhead compared to the basic ORAM.
The average performance loss is about 68.93\% and 75.10\%, respectively.
However, compared with the performance of Rcr-Baseline, the loss of Rcr-EHAP-ORAM is relatively small, about 3.65\%.
This is because recursive ORAM only needs to prevent the loss of data blocks in stash, so each time an ORAM access is performed, the corresponding backup data block of the accessed target block is persisted.

\begin{figure*}[htbp]
\subfigure[Performance comparison of EHAP-ORAM in 2-channel system.]
{
  \begin{minipage}{0.5\textwidth}
  \centering
  \includegraphics[width=0.95\textwidth ]{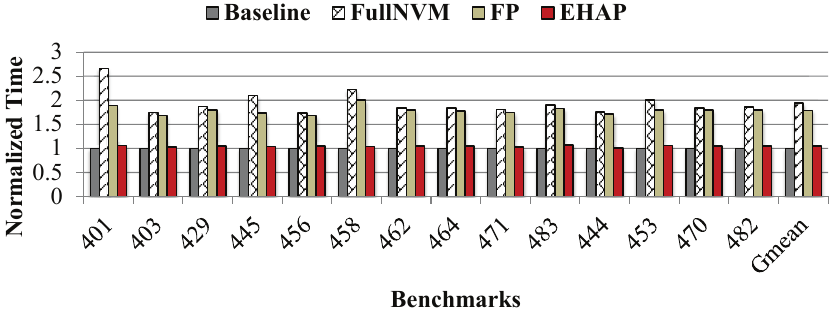}
  \label{2channelZ4PSORAMsystemPerformance}
  \end{minipage}
}
\subfigure[Performance comparison of EHAP-Recursive-ORAM in 2-channel system.]
{
  \begin{minipage}{0.5\textwidth}
  \centering
  \includegraphics[width=0.95\textwidth]{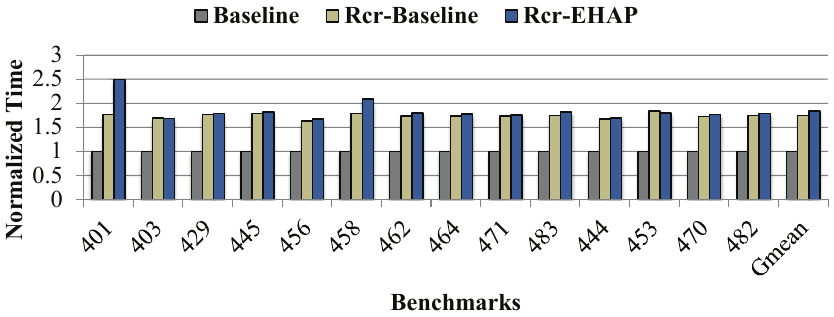}
  \label{2channelZ4RecursiveORAMsystemPerformance}
  \end{minipage}
  }
 \subfigure[Performance comparison of EHAP-ORAM in 4-channel system.]
{
  \begin{minipage}{0.5\textwidth}
  \centering
  \includegraphics[width=0.95\textwidth]{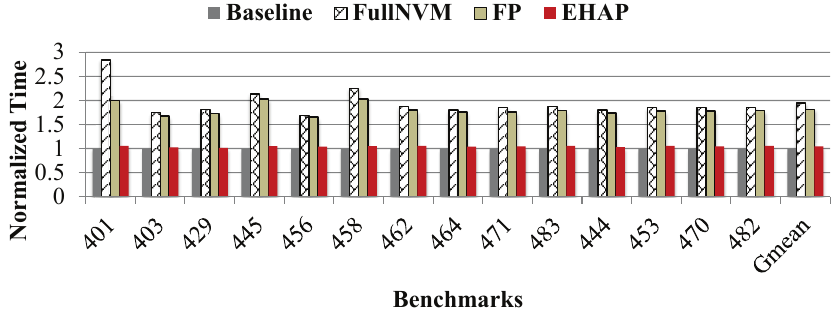}
  \label{4channelZ4PSORAMsystemPerformance}
  \end{minipage}
}
\subfigure[Performance comparison of EHAP-Recursive-ORAM in 4-channel system.]
{
  \begin{minipage}{0.5\textwidth}
  \centering
  \includegraphics[width=0.95\textwidth]{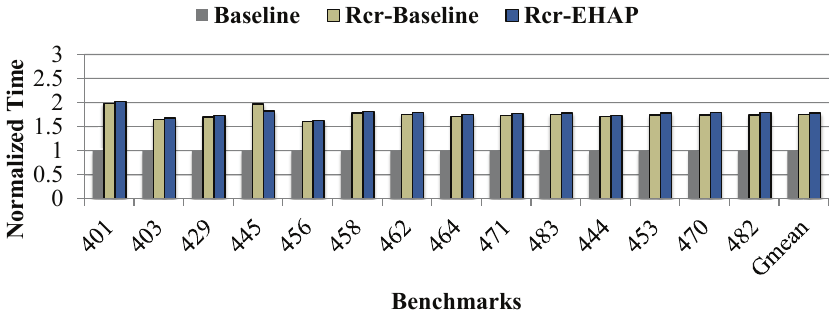}
  \label{4channelZ4RecursiveORAMsystemPerformance}
  \end{minipage}
  }
\vspace{-0.8em}
\caption{Multi-channel performance comparison.}
\label{Multi-channel-Z4}
\end{figure*}

\textbf{B. Single-Channel NVM read/write traffic.}
Figure \ref{1channel-NVMstash-RecursiveORAM-NVMORAM-AccessNumber}  shows the comparison of memory read/write traffic between an ORAM system without a persistent design and an ORAM system with a persistent design.
From Figure \ref{NumberofNormalizedReads}, we can see that when recursive ORAM executes ORAM read access, the number of read accesses increases significantly, the average increase was about 92.12\% and 93.25\%, respectively.
The number of other designs ORAM read accesses remains basically unchanged.
This is because recursive ORAM also performs ORAM access every time it accesses PosMap, resulting in a significant increase in reading traffic accesses.
The increase in the number of ORAM read accesses is one of the factors that cause the performance of the ORAM system to decrease.

The FullNVM design has the largest persistent write traffic overhead in Figure \ref{NumberofNormalizedWrites}, which increased by about 111.63\%.
Since each ORAM access requires the data to be unloaded from the NVM-ORAM tree onto the chip side of the NVM-stash, this will directly result in the addition of a large number of persistent writes.
The real blocks in the ORAM tree only account for 50\% of the total capacity of the ORAM tree.
Therefore, each time a data block is unloaded from the NVM-ORAM tree to the on-chip stash, the number of persistent writes is about 48 times (average times, cache line size $= 64$B, $L=23$, $Z=4$).
Must access NVM-stash when performing ORAM access, and due to the current limitations of device engineering materials, the speed of read/write access to NVM will be slower, resulting in greater performance overhead.
This results in high memory read/write traffic, which negatively impacts NVM lifetime.

Several other design designs have shown similar memory read/write traffic.
The EHAP-ORAM design has the least increase in write traffic, with an average of about 4.84\%.
Compared with FullNVM and FP, the write traffic of EHAP-ORAM decreased by 106.79\% and 96.07\%, respectively.
Because in EHAP-ORAM design, only metadata in PosMap corresponding to the target block in the path of the NVM-ORAM tree is persisted every time, and it is persisted in batches epoch.
Compared with the FP design, the EHAP-ORAM design reduces many redundant persistency operations of PosMap metadata.
Compared with the Rcr-Baseline and the Rcr-EHAP-ORAM design, the write traffic of the Rcr-EHAP-ORAM design increases, about 15.54\% , which is caused by the fact that the Rcr-EHAP-ORAM design needs to back up the accessed target data blocks every time the execution is a stash eviction.

\textbf{C. Multi-Channel Performance.}
This experiment evaluates different designs in a multi-channel system, where the workload is the same as a single-channel system.
Figure \ref{Multi-channel-Z4} shows the comparison of performance improvements for different channels.
From Figure \ref{2channelZ4PSORAMsystemPerformance}, Baseline indicates the performance of basic ORAM without a persistency in a 2-channel system, a similar baseline is in a 4-channel system.

In Figure \ref{2channelZ4PSORAMsystemPerformance}, most of the ORAM performance of the persistency design in the 2-channel system is better than the performance of the basic ORAM in the single-channel systems.
The performance improvement of the EHAP-ORAM design is close to Baseline, which shows the availability of the design the best.
Compared with Baseline, several persistency designs (FullNVM, FP and EHAP-ORAM) have an average performance reduction of 94.21\%, 78.99\%, and 4.66\%, respectively.

Similarly,  recursive ORAM systems with persistent design, as shown in Figure \ref{2channelZ4RecursiveORAMsystemPerformance}, in a 2-channel system, the performance improvement of the Rcr-EHAP-ORAM design approaches Rcr-Baseline.
Rcr-Baseline and Rcr-EHAP-ORAM compared to Baseline, the average performance decreased by 74.28\% and 82.14\%.
Rcr-EHAP-ORAM compared to Baseline, performance is reduced by about 4.51\%.

 When the number of channels is increased to 4, the performance of each persistency is not significantly improved compared to the performance of 2-channel, but there is still an improvement, as shown in Figure \ref{4channelZ4PSORAMsystemPerformance} and \ref{4channelZ4RecursiveORAMsystemPerformance}.
In the 4-channel system, the performance of FullNVM, FP and EHAP-ORAM compared to Baseline is decreased by 94.74\%, 80.91\%, and 4.60\%, respectively.
The performance of the Rcr-EHAP-ORAM is only 1.33\% lower than that of the Rcr-Baseline.

However, it is feasible to improve the performance of the EHAP-ORAM system only by adding channels, but the performance improvement is limited.
Therefore, in future work, to better the performance of the EHAP-ORAM system, we will continue to explore other performance optimization methods.


\section{Conclusions}
In this paper, we introduce a EHAP-ORAM system, a design designed to solve the previously proposed problem of ORAM system crash consistency.
To the best of our knowledge, this is the first work to solve the crash consistency problem of the ORAM system.
This requirement ensures that the memory persistency data and related metadata are recovered in a consistent state between system failures.

We first analyze the basic ORAM protocol without the persistency.
We find that data loss is easy to be caused if the system crashes or power failure when performing ORAM access, and the lost data cannot be effectively recovered, which eventually leads to the error of ORAM access.
To overcome the problem of crash consistency and effectively recover lost data, we have proposed several different persistency designs,  and low overhead hardware design is proposed to implement atomicity in the NVM-ORAM system with a crash consistency when performing ORAM access.
The experimental results show that the proposed persistency design is not only applicable to traditional ORAM systems, but also to recursive ORAM systems.
We believe that our work provides holistic system support for data persistency and security.

\bibliographystyle{plain}
\bibliography{OramFirstPaperRef,ref}

\begin{thebibliography}{10}

\bibitem{3dxpoint}
Intel optane technology: Revolutionizing memory and storage.
\newblock
  \url{https://www.intel.com/content/www/us/en/architecture-and-technology/intel-optane-technology.html}.
\newblock Accessed: 2020-03-30.

\bibitem{ahmad2019obfuscuro}
Adil Ahmad, Byunggill Joe, Yuan Xiao, Yinqian Zhang, Insik Shin, and
  Byoungyoung Lee.
\newblock Obfuscuro: A commodity obfuscation engine on intel sgx.
\newblock In {\em NDSS}, 2019.

\bibitem{ahmad2018obliviate}
Adil Ahmad, Kyungtae Kim, Muhammad~Ihsanulhaq Sarfaraz, and Byoungyoung Lee.
\newblock Obliviate: A data oblivious filesystem for intel sgx.
\newblock In {\em NDSS}, 2018.

\bibitem{arulraj2017build}
Joy Arulraj and Andrew Pavlo.
\newblock How to build a non-volatile memory database management system.
\newblock In {\em Proceedings of the 2017 ACM International Conference on
  Management of Data}, pages 1753--1758, 2017.

\bibitem{arulraj2015let}
Joy Arulraj, Andrew Pavlo, and Subramanya~R Dulloor.
\newblock Let's talk about storage \& recovery methods for non-volatile memory
  database systems.
\newblock In {\em Proceedings of the 2015 ACM SIGMOD International Conference
  on Management of Data}, pages 707--722, 2015.

\bibitem{awad2017obfus}
Amro Awad, Yipeng Wang, Deborah Shands, and Yan Solihin.
\newblock Obfusmem: A low-overhead access obfuscation for trusted memories.
\newblock In {\em Proceedings of the 44th Annual International Symposium on
  Computer Architecture}, pages 107--119, 2017.

\bibitem{awad2019triad}
Amro Awad, Mao Ye, Yan Solihin, Laurent Njilla, and Kazi~Abu Zubair.
\newblock Triad-nvm: Persistency for integrity-protected and encrypted
  non-volatile memories.
\newblock In {\em Proceedings of the 46th International Symposium on Computer
  Architecture}, pages 104--115, 2019.

\bibitem{bajikar2002trusted}
Sundeep Bajikar.
\newblock Trusted platform module (tpm) based security on notebook pcs-white
  paper.
\newblock {\em Mobile Platforms Group Intel Corporation}, 1:20, 2002.

\bibitem{benton2017ccix}
Brad Benton.
\newblock Ccix, gen-z, opencapi: Overview \& comparison.
\newblock In {\em 13th ANNUAL WORKSHOP 2017}, 2017.

\bibitem{binkert2011gem5}
Nathan Binkert, Bradford Beckmann, Gabriel Black, Steven~K Reinhardt, Ali
  Saidi, Arkaprava Basu, Joel Hestness, Derek~R Hower, Tushar Krishna, Somayeh
  Sardashti, et~al.
\newblock The gem5 simulator.
\newblock {\em ACM SIGARCH computer architecture news}, 39(2):1--7, 2011.

\bibitem{chen2013fsmac}
Jianxi Chen, Qingsong Wei, Cheng Chen, and Lingkun Wu.
\newblock Fsmac: A file system metadata accelerator with non-volatile memory.
\newblock In {\em 2013 IEEE 29th Symposium on Mass Storage Systems and
  Technologies (MSST)}, pages 1--11. IEEE, 2013.

\bibitem{choi201220nm}
Youngdon Choi, Ickhyun Song, Mu-Hui Park, Hoeju Chung, Sanghoan Chang,
  Beakhyoung Cho, Jinyoung Kim, Younghoon Oh, Duckmin Kwon, Jung Sunwoo, et~al.
\newblock A 20nm 1.8 v 8gb pram with 40mb/s program bandwidth.
\newblock In {\em 2012 IEEE International Solid-State Circuits Conference},
  pages 46--48. IEEE, 2012.

\bibitem{coburn2011nv}
Joel Coburn, Adrian~M Caulfield, Ameen Akel, Laura~M Grupp, Rajesh~K Gupta,
  Ranjit Jhala, and Steven Swanson.
\newblock Nv-heaps: making persistent objects fast and safe with
  next-generation, non-volatile memories.
\newblock {\em ACM SIGARCH Computer Architecture News}, 39(1):105--118, 2011.

\bibitem{condit2009better}
Jeremy Condit, Edmund~B Nightingale, Christopher Frost, Engin Ipek, Benjamin
  Lee, Doug Burger, and Derrick Coetzee.
\newblock Better i/o through byte-addressable, persistent memory.
\newblock In {\em Proceedings of the ACM SIGOPS 22nd symposium on Operating
  systems principles}, pages 133--146, 2009.

\bibitem{fletcher2015freecursive}
Christopher~W Fletcher, Ling Ren, Albert Kwon, Marten Van~Dijk, and Srinivas
  Devadas.
\newblock Freecursive oram: [nearly] free recursion and integrity verification
  for position-based oblivious ram.
\newblock In {\em Proceedings of the Twentieth International Conference on
  Architectural Support for Programming Languages and Operating Systems}, pages
  103--116, 2015.

\bibitem{fletcher2015low}
Christopher~W Fletcher, Ling Ren, Albert Kwon, Marten Van~Dijk, Emil Stefanov,
  Dimitrios Serpanos, and Srinivas Devadas.
\newblock A low-latency, low-area hardware oblivious ram controller.
\newblock In {\em 2015 IEEE 23rd Annual International Symposium on
  Field-Programmable Custom Computing Machines}, pages 215--222. IEEE, 2015.

\bibitem{goldreich1987towards}
Oded Goldreich.
\newblock Towards a theory of software protection and simulation by oblivious
  rams.
\newblock In {\em Proceedings of the nineteenth annual ACM symposium on Theory
  of computing}, pages 182--194, 1987.

\bibitem{goldreich1996software}
Oded Goldreich and Rafail Ostrovsky.
\newblock Software protection and simulation on oblivious rams.
\newblock {\em Journal of the ACM (JACM)}, 43(3):431--473, 1996.

\bibitem{hady2017platform}
Frank~T Hady, Annie Foong, Bryan Veal, and Dan Williams.
\newblock Platform storage performance with 3d xpoint technology.
\newblock {\em Proceedings of the IEEE}, 105(9):1822--1833, 2017.

\bibitem{henning2006spec}
John~L Henning.
\newblock Spec cpu2006 benchmark descriptions.
\newblock {\em ACM SIGARCH Computer Architecture News}, 34(4):1--17, 2006.

\bibitem{hu2020deepsniffer}
Xing Hu, Ling Liang, Shuangchen Li, Lei Deng, Pengfei Zuo, Yu~Ji, Xinfeng Xie,
  Yufei Ding, Chang Liu, Timothy Sherwood, et~al.
\newblock Deepsniffer: A dnn model extraction framework based on learning
  architectural hints.
\newblock In {\em Proceedings of the Twenty-Fifth International Conference on
  Architectural Support for Programming Languages and Operating Systems}, pages
  385--399, 2020.

\bibitem{islam2012access}
Mohammad~Saiful Islam, Mehmet Kuzu, and Murat Kantarcioglu.
\newblock Access pattern disclosure on searchable encryption: Ramification,
  attack and mitigation.
\newblock In {\em Ndss}, volume~20, page~12. Citeseer, 2012.

\bibitem{izraelevitz2019basic}
Joseph Izraelevitz, Jian Yang, Lu~Zhang, Juno Kim, Xiao Liu, Amirsaman
  Memaripour, Yun~Joon Soh, Zixuan Wang, Yi~Xu, Subramanya~R Dulloor, et~al.
\newblock Basic performance measurements of the intel optane dc persistent
  memory module.
\newblock {\em arXiv preprint arXiv:1903.05714}, 2019.

\bibitem{jiang2016crash}
Yanyan Jiang, Haicheng Chen, Feng Qin, Chang Xu, Xiaoxing Ma, and Jian Lu.
\newblock Crash consistency validation made easy.
\newblock In {\em Proceedings of the 2016 24th ACM SIGSOFT International
  Symposium on Foundations of Software Engineering}, pages 133--143, 2016.

\bibitem{johnson2016intel}
Simon Johnson, Vinnie Scarlata, Carlos Rozas, Ernie Brickell, and Frank Mckeen.
\newblock Intel{\textregistered} software guard extensions: Epid provisioning
  and attestation services.
\newblock {\em White Paper}, 1:1--10, 2016.

\bibitem{kaplan2016amd}
David Kaplan, Jeremy Powell, and Tom Woller.
\newblock Amd memory encryption.
\newblock {\em White paper}, 2016.

\bibitem{keeton2015machine}
Kimberly Keeton.
\newblock The machine: An architecture for memory-centric computing.
\newblock In {\em Workshop on Runtime and Operating Systems for Supercomputers
  (ROSS)}, volume~10, 2015.

\bibitem{komalan2018main}
Manu Komalan, Oh~Hyung Rock, Matthias Hartmann, Sushil Sakhare, Christian
  Tenllado, Jos{\'e}~Ignacio G{\'o}mez, Gouri~Sankar Kar, Arnaud Furnemont,
  Francky Catthoor, Sophiane Senni, et~al.
\newblock Main memory organization trade-offs with dram and stt-mram options
  based on gem5-nvmain simulation frameworks.
\newblock In {\em 2018 Design, Automation \& Test in Europe Conference \&
  Exhibition (DATE)}, pages 103--108. IEEE, 2018.

\bibitem{le2020tale}
Duc~V Le, Lizzy~Tengana Hurtado, Adil Ahmad, Mohsen Minaei, Byoungyoung Lee,
  and Aniket Kate.
\newblock A tale of two trees: One writes, and other reads: Optimized oblivious
  accesses to bitcoin and other utxo-based blockchains.
\newblock {\em Proceedings on Privacy Enhancing Technologies},
  2020(2):519--536, 2020.

\bibitem{lie2000architectural}
David Lie, Chandramohan Thekkath, Mark Mitchell, Patrick Lincoln, Dan Boneh,
  John Mitchell, and Mark Horowitz.
\newblock Architectural support for copy and tamper resistant software.
\newblock {\em Acm Sigplan Notices}, 35(11):168--177, 2000.

\bibitem{liu2018crash}
Sihang Liu, Aasheesh Kolli, Jinglei Ren, and Samira Khan.
\newblock Crash consistency in encrypted non-volatile main memory systems.
\newblock In {\em 2018 IEEE International Symposium on High Performance
  Computer Architecture (HPCA)}, pages 310--323. IEEE, 2018.

\bibitem{maas2013phantom}
Martin Maas, Eric Love, Emil Stefanov, Mohit Tiwari, Elaine Shi, Krste
  Asanovic, John Kubiatowicz, and Dawn Song.
\newblock Phantom: Practical oblivious computation in a secure processor.
\newblock In {\em Proceedings of the 2013 ACM SIGSAC conference on Computer \&
  communications security}, pages 311--324, 2013.

\bibitem{pillai2014all}
Thanumalayan~Sankaranarayana Pillai, Vijay Chidambaram, Ramnatthan Alagappan,
  Samer Al-Kiswany, Andrea~C Arpaci-Dusseau, and Remzi~H Arpaci-Dusseau.
\newblock All file systems are not created equal: On the complexity of crafting
  crash-consistent applications.
\newblock In {\em 11th $\{$USENIX$\}$ Symposium on Operating Systems Design and
  Implementation ($\{$OSDI$\}$ 14)}, pages 433--448, 2014.

\bibitem{trustzone}
Sandro Pinto and Nuno Santos.
\newblock Demystifying arm trustzone: A comprehensive survey.
\newblock {\em ACM Computing Surveys (CSUR)}, 51(6):1--36, 2019.

\bibitem{poremba2015nvmain}
Matthew Poremba, Tao Zhang, and Yuan Xie.
\newblock Nvmain 2.0: A user-friendly memory simulator to model (non-) volatile
  memory systems.
\newblock {\em IEEE Computer Architecture Letters}, 14(2):140--143, 2015.

\bibitem{rakshit2018leo}
Joydeep Rakshit and Kartik Mohanram.
\newblock Leo: Low overhead encryption oram for non-volatile memories.
\newblock {\em IEEE Computer Architecture Letters}, 17(2):100--104, 2018.

\bibitem{ren2015thynvm}
Jinglei Ren, Jishen Zhao, Samira Khan, Jongmoo Choi, Yongwei Wu, and Onur
  Mutiu.
\newblock Thynvm: Enabling software-transparent crash consistency in persistent
  memory systems.
\newblock In {\em 2015 48th Annual IEEE/ACM International Symposium on
  Microarchitecture (MICRO)}, pages 672--685. IEEE, 2015.

\bibitem{ren2015constants}
Ling Ren, Christopher~W Fletcher, Albert Kwon, Emil Stefanov, Elaine Shi,
  Marten Van~Dijk, and Srinivas Devadas.
\newblock Constants count: Practical improvements to oblivious ram.
\newblock In {\em USENIX Security Symposium}, pages 415--430, 2015.

\bibitem{ren2014unified}
Ling Ren, Christopher~W Fletcher, Xiangyao Yu, Albert Kwon, Marten van Dijk,
  and Srinivas Devadas.
\newblock Unified oblivious-ram: Improving recursive oram with locality and
  pseudorandomness.
\newblock {\em IACR Cryptol. ePrint Arch.}, 2014:205, 2014.

\bibitem{ren2013design}
Ling Ren, Xiangyao Yu, Christopher~W Fletcher, Marten Van~Dijk, and Srinivas
  Devadas.
\newblock Design space exploration and optimization of path oblivious ram in
  secure processors.
\newblock In {\em Proceedings of the 40th Annual International Symposium on
  Computer Architecture}, pages 571--582, 2013.

\bibitem{sahin2016taostore}
Cetin Sahin, Victor Zakhary, Amr El~Abbadi, Huijia Lin, and Stefano Tessaro.
\newblock Taostore: Overcoming asynchronicity in oblivious data storage.
\newblock In {\em 2016 IEEE Symposium on Security and Privacy (SP)}, pages
  198--217. IEEE, 2016.

\bibitem{sasy2017zerotrace}
Sajin Sasy, Sergey Gorbunov, and Christopher~W Fletcher.
\newblock Zerotrace: Oblivious memory primitives from intel sgx.
\newblock {\em IACR Cryptology ePrint Archive}, 2017:549, 2017.

\bibitem{sehgal2015empirical}
Priya Sehgal, Sourav Basu, Kiran Srinivasan, and Kaladhar Voruganti.
\newblock An empirical study of file systems on nvm.
\newblock In {\em 2015 31st Symposium on Mass Storage Systems and Technologies
  (MSST)}, pages 1--14. IEEE, 2015.

\bibitem{seshadri2013rowclone}
Vivek Seshadri, Yoongu Kim, Chris Fallin, Donghyuk Lee, Rachata
  Ausavarungnirun, Gennady Pekhimenko, Yixin Luo, Onur Mutlu, Phillip~B
  Gibbons, Michael~A Kozuch, et~al.
\newblock Rowclone: fast and energy-efficient in-dram bulk data copy and
  initialization.
\newblock In {\em Proceedings of the 46th Annual IEEE/ACM International
  Symposium on Microarchitecture}, pages 185--197, 2013.

\bibitem{seshadri2015page}
Vivek Seshadri, Gennady Pekhimenko, Olatunji Ruwase, Onur Mutlu, Phillip~B
  Gibbons, Michael~A Kozuch, Todd~C Mowry, and Trishul Chilimbi.
\newblock Page overlays: An enhanced virtual memory framework to enable
  fine-grained memory management.
\newblock {\em ACM SIGARCH Computer Architecture News}, 43(3S):79--91, 2015.

\bibitem{stefanov2013path}
Emil Stefanov, Marten Van~Dijk, Elaine Shi, Christopher Fletcher, Ling Ren,
  Xiangyao Yu, and Srinivas Devadas.
\newblock Path oram: an extremely simple oblivious ram protocol.
\newblock In {\em Proceedings of the 2013 ACM SIGSAC conference on Computer \&
  communications security}, pages 299--310, 2013.

\bibitem{swami2016secret}
Shivam Swami, Joydeep Rakshit, and Kartik Mohanram.
\newblock Secret: Smartly encrypted energy efficient non-volatile memories.
\newblock In {\em Proceedings of the 53rd Annual Design Automation Conference},
  pages 1--6, 2016.

\bibitem{venkataraman2011consistent}
Shivaram Venkataraman, Niraj Tolia, Parthasarathy Ranganathan, Roy~H Campbell,
  et~al.
\newblock Consistent and durable data structures for non-volatile
  byte-addressable memory.
\newblock In {\em FAST}, volume~11, pages 61--75, 2011.

\bibitem{volos2011mnemosyne}
Haris Volos, Andres~Jaan Tack, and Michael~M Swift.
\newblock Mnemosyne: Lightweight persistent memory.
\newblock {\em ACM SIGARCH Computer Architecture News}, 39(1):91--104, 2011.

\bibitem{wang2018d}
Rujia Wang, Youtao Zhang, and Jun Yang.
\newblock D-oram: Path-oram delegation for low execution interference on cloud
  servers with untrusted memory.
\newblock In {\em 2018 IEEE International Symposium on High Performance
  Computer Architecture (HPCA)}, pages 416--427. IEEE, 2018.

\bibitem{wu2016nvmcached}
Xingbo Wu, Fan Ni, Li~Zhang, Yandong Wang, Yufei Ren, Michel Hack, Zili Shao,
  and Song Jiang.
\newblock Nvmcached: An nvm-based key-value cache.
\newblock In {\em Proceedings of the 7th ACM SIGOPS Asia-Pacific Workshop on
  Systems}, pages 1--7, 2016.

\bibitem{yang2019no}
Fan Yang, Youyou Lu, Youmin Chen, Haiyu Mao, and Jiwu Shu.
\newblock No compromises: Secure nvm with crash consistency, write-efficiency
  and high-performance.
\newblock In {\em 2019 56th ACM/IEEE Design Automation Conference (DAC)}, pages
  1--6. IEEE, 2019.

\bibitem{yang2015nv}
Jun Yang, Qingsong Wei, Cheng Chen, Chundong Wang, Khai~Leong Yong, and
  Bingsheng He.
\newblock Nv-tree: Reducing consistency cost for nvm-based single level
  systems.
\newblock In {\em 13th $\{$USENIX$\}$ Conference on File and Storage
  Technologies ($\{$FAST$\}$ 15)}, pages 167--181, 2015.

\bibitem{young2015deuce}
Vinson Young, Prashant~J Nair, and Moinuddin~K Qureshi.
\newblock Deuce: Write-efficient encryption for non-volatile memories.
\newblock {\em ACM SIGARCH Computer Architecture News}, 43(1):33--44, 2015.

\bibitem{zhang2018shadow}
Xian Zhang, Guangyu Sun, Peichen Xie, Chao Zhang, Yannan Liu, Lingxiao Wei,
  Qiang Xu, and Chun~Jason Xue.
\newblock Shadow block: accelerating oram accesses with data duplication.
\newblock In {\em 2018 51st Annual IEEE/ACM International Symposium on
  Microarchitecture (MICRO)}, pages 961--973. IEEE, 2018.

\bibitem{zhang2015fork}
Xian Zhang, Guangyu Sun, Chao Zhang, Weiqi Zhang, Yun Liang, Tao Wang, Yiran
  Chen, and Jia Di.
\newblock Fork path: improving efficiency of oram by removing redundant memory
  accesses.
\newblock In {\em 2015 48th Annual IEEE/ACM International Symposium on
  Microarchitecture (MICRO)}, pages 102--114. IEEE, 2015.

\bibitem{zhuang2004hide}
Xiaotong Zhuang, Tao Zhang, and Santosh Pande.
\newblock Hide: an infrastructure for efficiently protecting information
  leakage on the address bus.
\newblock {\em ACM SIGOPS Operating Systems Review}, 38(5):72--84, 2004.

\bibitem{zuo2019supermem}
Pengfei Zuo, Yu~Hua, and Yuan Xie.
\newblock Supermem: Enabling application-transparent secure persistent memory
  with low overheads.
\newblock In {\em Proceedings of the 52nd Annual IEEE/ACM International
  Symposium on Microarchitecture}, pages 479--492, 2019.

\end{thebibliography}

\end{document}